# Semi-Classical Monte Carlo Simulation of Contact Geometry, Orientation, and Ideality on Nano-scale Si and III-V n-channel FinFETs in the Quasi-Ballistic Limit

Aqyan A. Bhatti,[1,a)] Dax M. Crum,[2,b)] Amith Valsaraj,[2,c)] Leonard F. Register[2], and Sanjay K. Banerjee[2]

[1]*Texas Materials Institute and the Materials Science and Engineering Program, The University of Texas at Austin,*

*Austin, Texas, USA*

[2]*Microelectronics Research Center, The University of Texas at Austin, Austin, Texas, USA*

The effects of contact geometry and ideality on InGaAs and Si nano-scale n-channel FinFET performance are studied using a quantum-corrected semi-classical Monte Carlo method. Illustrative end, saddle/slot, and raised source/drain contacts were modeled, and with ideal transmissivity and reduced transmissivity more consistent with experimental contact resistivities. Far-from-equilibrium degenerate statistics, quantum-confinement effects on carrier distributions in real-space and among energy valleys, quasi-ballistic transport inaccessible through drift-diffusion and hydrodynamic simulations, and scattering mechanisms and contact geometries not readily accessible through non-equilibrium Green's function simulation are addressed. Silicon ⟨110⟩ channel devices, Si ⟨100⟩ channel devices, multi-valley (MV) InGaAs devices with conventionally-reported energy valley offsets, and idealized Γ-valley only (Γ) InGaAs devices are modeled. Simulated silicon devices exhibited relatively limited degradation in performance due to non-ideal contact transmissivities, more limited sensitivity to contact geometry with non-ideal contact transmissivities, and some contact-related advantage for Si ⟨110⟩ channel devices. In contrast, simulated InGaAs devices were highly sensitive to contact geometry and ideality and the peripheral valley's energy offset. It is illustrative of this latter sensitivity that simulated Γ-InGaAs device outperformed all others by a factor of two or more in terms of peak transconductance with perfectly transmitting reference end contacts, while silicon devices outperformed Γ-InGaAs for all contact geometries with non-ideal transmissivities, and MV-InGaAs devices performed the poorest under all simulation scenarios. While these specific results are not encouraging for InGaAs devices, they may help guide the design of high-mobility channel devices through the identification of performance bottlenecks.

___________________________

a) Author to whom correspondence should be addressed. Electronic mail: abhatti@utexas.edu.

b) D. M. Crum was with the Microelectronics Research Center, The University of Texas at Austin, Austin, USA at the time of his contribution to this work. He is now at Intel Corporation, USA.

c) A. Valsaraj was with the Microelectronics Research Center, The University of Texas at Austin, Austin, USA at the time of his contribution to this work. He is now at GlobalFoundries USA Inc., USA.

**I. INTRODUCTION**

New materials and new device designs continue to emerge toward extending CMOS scaling, including the possible use of high mobility and thermal velocity channel materials [1]. In direct gap III-V materials, Γ-valley conduction band electron mobilities and thermal velocities can be much greater than in silicon. Moreover, substantial ballistic transport can occur on scales greater than 100 nm [2] vs. on the scale of 10s of nm for Si (based on average velocity magnitude and scattering rate for thermal electrons). Full-quantum transport simulations of nanowire n-MOSFETs have indicated that III-Vs could compete with and outperform Si in high-performance (high $V_{DD}$, high $I_{off}$) applications [3]. However, with quasi-ballistic transport, the inability of the source region to replenish carriers in the channel—"source starvation"—also can become important with realistic contact geometries [4]. Moreover, with strong quantum confinement effects within the channel for low-effective-mass electrons and uncertainties about peripheral valley energy offsets, [5–9], the ability to take advantage of the performance advantages offered by Γ-valley electrons is not a given.

For this work, InGaAs and Si nano-scale n-channel FinFET performance as a function of contact geometry and ideality and peripheral valley energy in the former, and channel orientation in the latter, has been studied using a quantum-corrected semi-classical Monte Carlo method. Illustrative end, saddle/slot, and raised source/drain contacts were modeled, and each with both ideal transmissivity and reduced transmissivity more consistent with experimental contact resistivities. Far-from-equilibrium degenerate statistics, quantum-confinement effects on carrier distributions in real-space and among energy valleys, quasi-ballistic transport inaccessible through drift-diffusion and hydrodynamic simulations, and scattering mechanisms (including polar-optical phonon scattering in III-Vs) and contact geometries not readily accessible through non-equilibrium Green's function (NEGF) simulations are addressed. Silicon ⟨110⟩ channel devices, Si ⟨100⟩ channel devices, multi-valley (MV) InGaAs devices with conventionally-reported energy valley offsets, and idealized Γ-valley only (Γ) InGaAs devices are modeled.

Among our findings, simulated silicon devices exhibited relatively limited degradation in performance due to non-ideal contact transmissivities (within the range considered), more limited sensitivity to contact geometry with non-ideal contact transmissivities, and some contact-related advantage for Si ⟨110⟩ channel devices. In contrast, simulated InGaAs devices were highly sensitive to contact geometry and ideality and to the peripheral valley's energy offset. For example, despite challenges to simulation-based comparison with different channel materials, it is illustrative of this latter sensitivity that simulated Γ-InGaAs device outperformed all others by a factor of two or more in terms of peak transconductance with perfectly transmitting reference end contacts, although significantly less so in terms of on-state current above a constant current threshold due to a poorer turn on characteristic, while silicon devices outperformed Γ-InGaAs for all contact geometries with non-ideal transmissivities, and MV-InGaAs devices performed the poorest under all simulation scenarios.

The results here for common contact geometries, reasonable contact resistivities, and conventional expectations for peripheral valley energy offsets are not encouraging. However, the substantial advantage of the simulated ideal end-contact Γ-InGaAs device, at least in transconductance, continues to bear out the motivation for the use of such high thermal velocity materials. Perhaps the juxtaposition of these results can help identify the performance bottlenecks and, thus, opportunities for improvement.

## A. On contacts and InGaAs band structure

Common contact options for making contacts to multi-gate MOSFET/FinFET geometries include: 1) dumbbell-shaped source and drain contacts, 2) saddle or slot contacts, and 3) raised source-drain contacts (RSD) on merged fins. The dumbbell source and drain contact layout is like that of planar MOSFET contact in that contact holes (vias) are etched using a contact window mask down to the surface to be contacted. However, dumbbell layouts are not area efficient, and FinFETs are moving toward pad-less fin structures, such as saddle contacts or contacts to epitaxially-thickened S/D regions [10,11]. Saddle contacts are attractive because of a significantly smaller device footprint than the dumbbell layout, and the saddle metal contact couples to the fin top and sidewall surfaces through a thin metal silicide interface, potentially giving rise to a larger contact area to reduce contact resistance. If making a simple saddle contact to individual fins is not possible due to tight alignment tolerances, slot contacts, a variant on the theme, can be used instead, where a thicker layer of metal silicide is deposited across the S/D of all the fins, followed by metal contact across the top of the silicide as a whole. However, the extra contact metal between the fins in the latter case increases the parasitic gate-to-contact capacitance, which can limit circuit performance. A more attractive option is to increase the fin width in the S/D semiconductor regions by epitaxial growth, even to the point of merging adjacent fins (although not modeled as such here), in the RSD structure to eliminate the contact-to-fin pitch matching requirements and increase the surface area of the contact. In addition, the RSD structure has been shown to reduce parasitic S/D resistance and capacitance, but not at the expense of fin pitch [12–14]. One drawback of the RSD approach is the conformation of the source and drain surface depends on the source and drain epitaxial faceting. For (110) sidewalls, the final fin cross section is hexagonal or diamond-shaped, and hence, the contact will land on a non-planar surface. For (100) sidewalls, the cross section of the epitaxially grown semiconductor is rectangular and contacts will land on a flat surface. In any case, a common trait of these contacts relevant to this work is that each may be considered as "side" or "wrapped" contacts with respect to the channel orientation, in contrast to the end contacts that also are considered in this work for reference.

Parasitic source/drain series resistance $R_{series}$ also has been shown to play an increasingly limiting role in the performance of devices near the end of the International Technology Roadmap for Semiconductors (ITRS) [15]. $R_{series}$ can be divided into the four components: (1) extension-to-gate overlap resistance ($R_{OV}$), (2) S/D extension resistance ($R_{EXT}$), (3) deep S/D resistance ($R_{S/D}$), and (4) contact resistance between the semi-metallic silicide and the heavily doped semiconducting S/D interface ($R_C$). Historically, $R_{series}$ had been kept to about 10% of the intrinsic (ideal) channel saturation resistance, $R_{on}$; however, for nodes since 2008, $R_{series}$ has been approximately 25% of $R_{on}$. Moreover, continually decreasing device sizes have increased the contribution of $R_C$ to $R_{series}$, already contributing about 40% of $R_{series}$ at 50 nm gate lengths [16].

With regard to the considered InGaAs system, we note that the inter-valley separations between the light-mass Γ-valley and heavy-mass peripheral L-valleys ($\Delta E_{\Gamma\text{-L}}$) and X-valleys ($\Delta E_{\Gamma\text{-X}}$) are not reliably known, with significant uncertainty in $\Delta E_{\Gamma\text{-L}}$ in the literature [8]. A commonly cited tight-binding calculation places $\Delta E_{\Gamma\text{-L}}$ at 460 meV [6], while the only experimental determination places $\Delta E_{\Gamma\text{-L}}$ at 550 meV [6,7]. Recent density-functional calculations have even estimated $E_{GL}$ to be as large as 1.31 eV [9].

## II. SIMULATED FINFET STRUCTURES AND SIMULATION METHOD

### A. Device structure

For this paper, we investigate the effect of model saddle and RSD contacts on nano-scale Si and InGaAs n-channel FinFETs. We also investigate the performance of FinFETs with a reference S/D end-contact geometry, representing simpler contact models, including those used commonly in non-equilibrium Green's function (NEGF) simulations.

We model 18 nm gate length ($L_G$) and 6 nm fin width ($W_{FIN}$) In$_{0.53}$Ga$_{0.47}$As (which, henceforth, we will refer to simply as "InGaAs") and Si-channel devices, with illustrative end, saddle/slot [17], and RSD contact geometries, as shown in Fig. 1(a)-(c), respectively, with device geometry parameters listed in Table I. The source (drain) contact area for these device geometries are 228 nm$^2$, 574 nm$^2$ and 752 nm$^2$, respectively. The model saddle/slot and RSD geometries are intended to be illustrative of common contact geometries as discussed above; the end contact geometry represents one limit of performance and a geometry commonly used in NEGF simulations. The contact resistivities are varied from the ideal Landauer-Büttiker limit [18,19] to more realistic values by varying the transmission probability across the contact surface, as detailed elsewhere.

A ⟨100⟩ substrate orientation is considered for all devices. For Si, we considered both ⟨110⟩ and ⟨100⟩ channel orientations, which, among other things, leads to different degrees of quantum confinement normal to the channel orientation within the channel and drain extensions for electrons within the various otherwise-equivalent band-edge Δ-valleys. However, for the RSD contact geometry, for both Si channel orientations, we used the same rectangular geometry characteristic of ⟨100⟩ channel orientations. For InGaAs devices, with the band edge at the Γ-point, we considered only the ⟨100⟩ channel orientation. We assume a 3 nm thick HfO$_2$ gate oxide for all devices for the purpose of modeling electrostatic gate control. However, we also include a commonly-occurring thin interfacial SiO$_2$ layer in the silicon devices only for modeling the oxide effective mass in the quantum-confinement calculations, $m_{ox}^* = 0.55 m_e$ of SiO$_2$ vs. $m_{ox}^* = 0.15 m_e$ of HfO$_2$, where $m_e$ is the free electron mass [20,21]. The fin height ($H_{FIN}$) and oxide substrate thickness ($H_{BOX}$) of all devices are 35 nm and 10 nm, respectively. The source/drain regions, located 5 nm away from the edge of the gate region, are uniformly doped to 2×10$^{20}$ cm$^{-3}$ for silicon, and 5×10$^{19}$ cm$^{-3}$ for InGaAs, the maximum experimentally observed electrically active dopant concentrations of arsenic in silicon, and of silicon in $In_{0.53}Ga_{0.47}As$, respectively [22–24]. The devices have a decade/nm doping profile in the 5 nm source and drain extensions ($L_{EXT}$).

(a) end injection

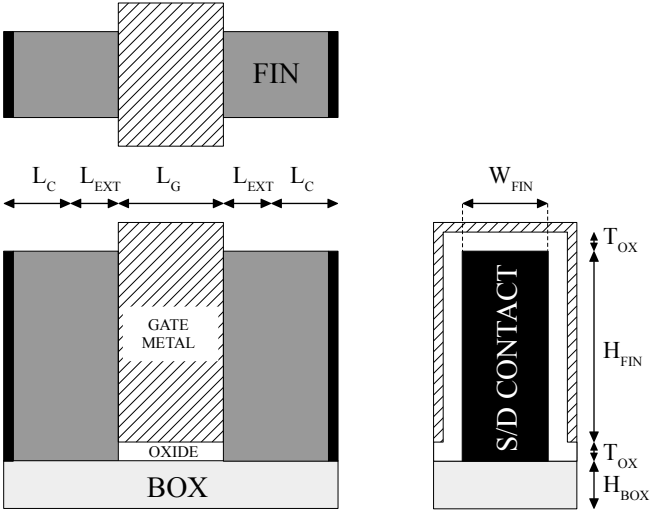

(b) saddle/slot

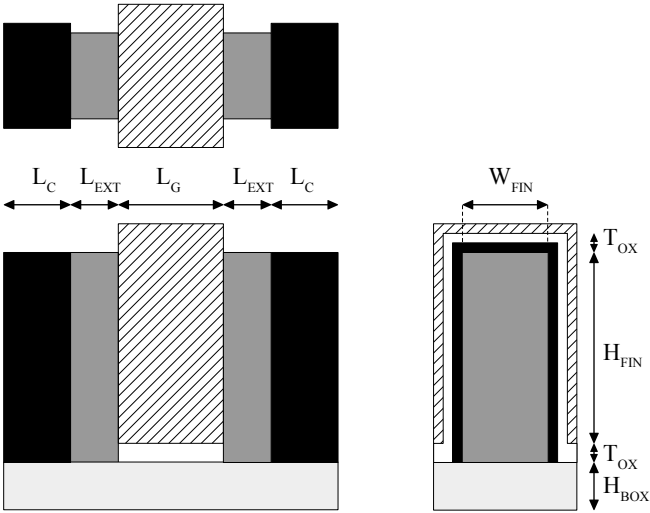

(c) raised source and drain

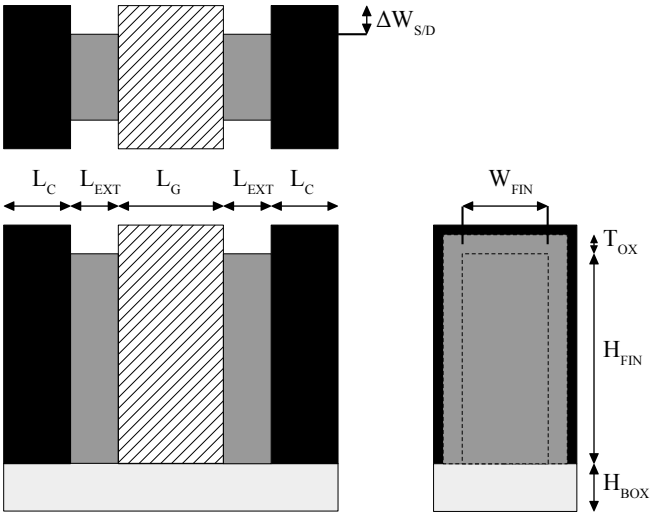

**FIG. 1.** Schematics of the simulated modeled FinFET geometries with (a) reference end contacts, (b) saddle/slot contacts, and (c) raised source and drain (RSD) geometries. For each, a side view (lower left), a top view (top), and an end view (right) are shown. The spacer regions are not shown in order to show the underlying semiconductor fin, shaded in grey. The hatched region represents the gate metal. The gate oxide located underneath the gate metal is visible in the end views of end and saddle/slot contact model devices. The source and drain contact surfaces are shown in black. We note that for the saddle/slot geometry, the source and drain contacts extend further to the side and above than shown, to the edge of the simulation region; however, only the near-source/drain-surface portions are shown for visual clarity. The source (drain) contact area for these device geometries are 228 nm$^2$, 574 nm$^2$, and 752 nm$^2$, respectively.

TABLE I. Device dimensions for the modeled FinFETs.

| Dimension | End, Saddle/Slot | Raised Source/Drain |
|---|---|---|
| $L_c$ | 8 nm | 8 nm |
| $L_{EXT}$ | 5 nm | 5 nm |
| $L_G$ | 18 nm | 18 nm |
| $H_{FIN}$ | 35 nm | 35 nm |
| $W_{FIN}$ | 6 nm | 6 nm |
| $H_{BOX}$ | 10 nm | 10 nm |
| $T_{OX}$ | 3 nm | 3 nm |
| $\Delta W_{S/D}$ | 0 nm | 6 nm |

### B. InGaAs band structure models

We considered two models of the In$_{0.53}$Ga$_{0.47}$As band structure, the MV-InGaAs model and Γ-model, as noted. In our MV-InGaAs model, we took the inter-valley separation between the light-mass Γ-valley and heavy-mass peripheral L-valleys and X-valleys as $\Delta E_{\Gamma\text{-}L}$ = 487 meV and $\Delta E_{\Gamma\text{-}X}$ = 610 meV, respectively, as determined by a set of bowing parameters recommended by Vurgaftman and colleagues in their comprehensive review article [5], which lies between the above-noted tight-binding and experimental values of [6,7], respectively. With the assumed 5×10$^{19}$ cm$^{-3}$ doping for InGaAs, the equilibrium Fermi energy is found nearly 500 meV above the conduction band edge, high enough to place approximately 40% of the equilibrium bulk carrier concentration in the L-valleys for the assumed Γ-to-L energy valley separation, as a consequence of the degenerate statistics, the much larger L valley than Γ-valley mass, and four-fold L-valley degeneracy [25]. In contrast, in Si the Fermi energy is found only approximately 100 meV above the conduction band edge with the degenerate statistics, despite the four-fold larger assumed doping. The Γ-InGaAs model ($\Delta E_{\Gamma\text{-}L} \to \infty$, $\Delta E_{\Gamma\text{-}X} \to \infty$) represents an ideal limiting behavior, but also is effectively consistent with the $\Delta E_{\Gamma\text{-}L}$ = 1.31 eV result of [9] given the maximum 0.6 V drain voltage and still smaller gate overdrive voltages considered here. Note that this work is moot on which model of InGaAs is more appropriate, Γ-InGaAs or MV-InGaAs; we simply consider the consequences of both.

We would be remiss not to note that at degenerate doping levels, charge carriers are not created by ionization of donor states to the conduction band with a commensurate rise in the Fermi level, but by merging the donor states with energy valley edges and a commensurate lowering of the conduction band edge below the Fermi level [26,27]. In this way, in particular, the effective peripheral valley separations in MV-InGaAs in the source and drain would be larger that otherwise expected, and the peripheral valley occupations would be reduced or eliminated, accordingly. This physics is not addressed in the simple band-structure models of this work. However, as discussed later, the modeled specific contact resistivities, ideal and non-ideal, of InGaAs are only weakly dependent on the assumed energy valley separations. Within the undoped channel region, energy valley separations are not thus impacted by the doping, while actually being reduced considerably by quantum confinement. And, although there may be some advantage to reducing the fraction of carriers in the peripheral valleys in the source (and drain), we found previously [25] that the peripheral valleys in the channel become heavily occupied in the ON-state in modeled MV-InGaAs device through inter-valley scattering even when not occupied in the modeled source and drain under lower doping (which, in turn, becomes a concern for any fully-ballistic simulations of multi-valley direct gap materials).

## C. Simulation methodology (essential elements)

We employed our in-house quantum-corrected three-dimensional semi-classical Monte Carlo (SCMC) methodology, University of Texas Monte Carlo (UTMC) [25], to study contact geometry and crystal orientation effects on carrier injection in Si and InGaAs n-channel FinFETs, while also modeling far-from-equilibrium degenerate statistics, non-ideal contacts, and quantum-confinement effects on carrier distributions in real-space and among energy valleys, and on phonon, impurity, and surface roughness scattering.

UTMC models carrier transport within 3-D device geometries considering intra- and inter-valley phonon (acoustic, deformation potential optical, polar optical and intervalley), surface roughness (SR), alloy, and ionized impurity scattering. Following the approach of Jacoboni and Fischetti, the electron energy bands are modeled analytically with non-parabolicity corrections [2,28], reasonable for the limited carrier energies considered here. Deformation potentials were adjusted to reproduce available experimental bulk mobility data. For ⟨100⟩ Si, surface roughness parameters then were adjusted to reproduce available interface mobility data, and the same surface roughness parameters then are used for Si ⟨110⟩ and InGaAs devices, which is likely optimistic, at least for now [29].

Because of high doping concentrations, degenerate statistics must be addressed. However, because of the far-from-equilibrium conditions encountered in these devices, carrier statistics cannot be approximated accurately using Fermi-Dirac distributions. Instead, UTMC directly models Pauli-Blocking (PB) of scattering to obtain the far-from-equilibrium local electron occupation probabilities from the local electron populations, $N(r, E, g, \pm)$, as a function of position ($r$), energy valley ($g$) and energy ($E$), and propagation direction, forward toward the drain end ($+$) or backward toward the source end ($-$).

Position-dependent and energy-valley-dependent quantum-corrected potentials (QCPs) are calculated to match the calculated quantum-corrected—as an approximation, for computational efficiency, for this purpose only—equilibrium classical carrier distributions to the quantum mechanical ones. The latter are obtained via self-consistent coupling of Schrödinger's time-independent equation with the Poisson's equation, while allowing for barrier penetration effects, which can moderate the effects of confinement significantly. For practicality, the QCPs are computed within 2-D cross sections normal to the channel direction within an effective mass approximation with a non-parabolicity correction. To approximate 3-D effects, the quantum corrections are ramped on starting at the onset of confinement at the source/drain-to-source/drain-extension boundaries, over a distance equal to the actual channel width. The quantum corrections then serve to: increase thresholds and alter relative valley occupancy, redistribute the carriers in real-space away from potential barriers, generally increase even intra-valley phonon scattering rates, particularly for randomizing processes, and determine the surface roughness scattering rate. In this latter way, although the employed surface roughness parameters for all devices here are taken as the same, the resulting surface roughness scattering is not.

Of particular note for this work, the contacts are modeled as in equilibrium in these UTMC simulations, with, for ideal contacts, electrons injected from the contacts into the S/D according to a velocity-weighted Fermi-Dirac distribution, while electrons reaching the contact surface from the source/drain regions are perfectly absorbed, such that under overall equilibrium conditions, the net current through the contacts vanishes on average even while electrons continue to be injected and absorbed. Details of the models and methodology can be found in [25]. Non-ideal contacts are then considered by scaling down the electron injection and absorption probabilities across the contact interfaces consistent with experimental results, as discussed further subsequently.

### III. RESULTS AND DISCUSSION

#### A. Common performance measures and results for ideal contacts

To analyze device performance, we initially compare peak transconductance ($g_m$), on-current ($I_{on}$), subthreshold swing ($S$), and drain-induced barrier lowering (DIBL) in the off-state, as well as the turn-on abruptness, as measured by the difference ($\Delta V_T$) between two different estimates of threshold voltage ($V_T$). $g_m = \max(\partial I_{DS}/\partial V_{GS})$, where $I_{DS}$ and $V_{GS}$ are the drain-to-source current and gate-to-source voltage, respectively, is obtained from a centered moving average over an interval of ten gate-to-source voltage samples to reduce noise in the data. The drain-source voltage $V_{DS}$ in these calculations is set to the supply voltage $V_{DD}$ in this section, which, in turn, is taken to be 0.6 V, in accordance with ITRS predictions for future scaled MOSFETS [15]. $V_{GS}$ then was swept from OFF to ON in steps of 25 mV (and somewhat beyond the 0.6 V range in practice to allow for initially unknown thresholds and exhibition of some behavior beyond the normal operating regime). The drain current was divided by the fin perimeter ($2H_{FIN} + W_{FIN}$ in Fig. 1) for the purpose of calculating current density. The turn-on abruptness measure is, specifically, $\Delta V_T = V_T^{ELR} - V_T^{CC}$. Here, $V_T^{CC}$ is the threshold voltage as obtained by the constant current (CC) method, which is widely used in industry and serves as a reference for our $I_{on}$ calculations, where the threshold is defined by a fixed $I_{DS}$ target. In this work, we take $I_{DS}(V_T^{CC}) = 0.01$ mA/μm, which, although larger than values typically used experimentally, serves our purpose sufficiently while maintaining satisfactory statistics. $V_T^{ELR}$ is the threshold voltage as obtained by extrapolation in the linear region (ELR) [30], where $V_T$ is defined by linear extrapolation from the point of maximum slope (peak transconductance) of the $I_{DS}$ vs. $V_{GS}$ curve in the ON-state, back to the intercept with the gate-to-source voltage ($V_{GS}$) axis. The ON-state current, $I_{on}$, is then calculated at the gate overdrive above threshold $V_{on} = V_{GS} - V_T^{CC} = 0.35$ V with, again, $V_{DS} = 0.6$ V. DIBL $= -d\Phi_b/dV_{DS}$, where $\Phi_b$ is the channel potential barrier, is calculated below threshold with $V_{DS} = 0.6$ V. Thus, the reported ON-state currents are dependent on the values of both $g_m$ and $\Delta V_T$. Subthreshold swing, $S = (\ln 10) dV_{GS}/d(\ln I_{DS})$, is calculated in terms of $\Phi_b$ within a simple thermionic emission model due to the lack of sufficient statistics for direct calculation with small currents well-below threshold, and under zero $V_{DS}$, representing the linear regime of operation. Simulation results are provided in Figs. 2 and 3 and discussed in detail below.

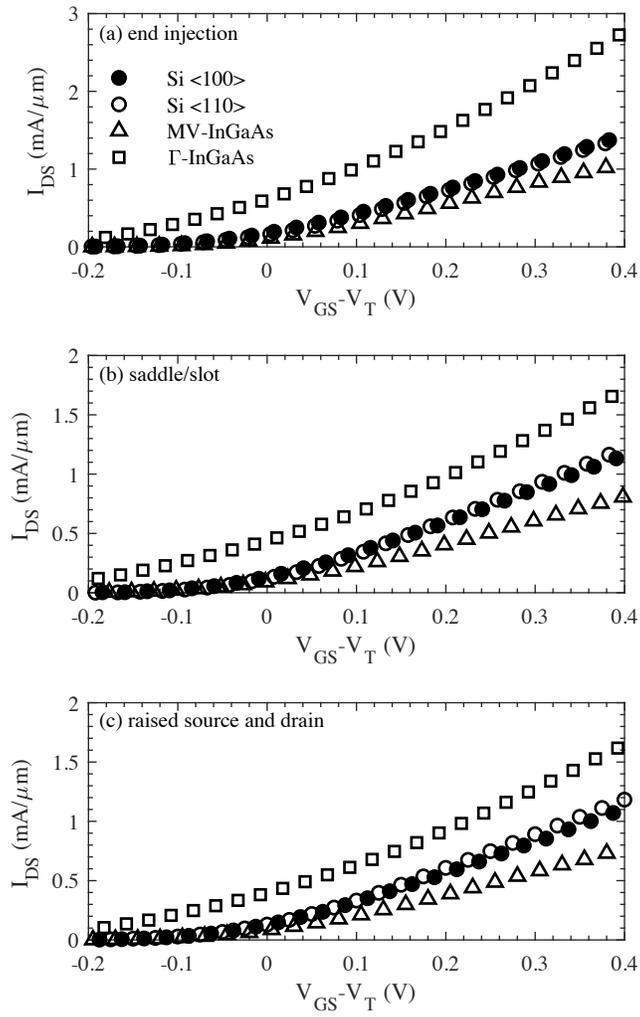

**FIG. 2.** $I_{DS}$-$V_{GS}$ simulation results for 18 nm channel length Si ⟨110⟩ (open circles), Si ⟨100⟩ (solid circles), MV-In$_{0.53}$Ga$_{0.47}$As (open triangles), and Γ-In$_{0.53}$Ga$_{0.47}$As (open squares) FinFETs for (a) end injection, (b) saddle/slot, and (c) raised source and drain. $V_{DS} = 0.6$ V. For visual clarity, the threshold voltage is that obtained using the extrapolation in the linear regime method.

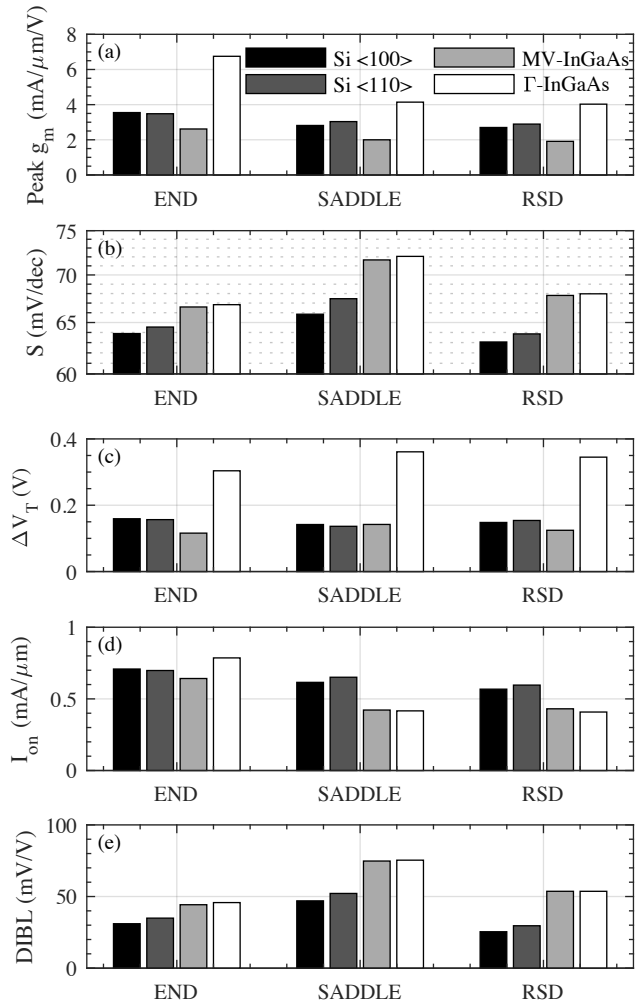

**FIG. 3.** Dependence of (a) (centered moving average of) the peak transconductance $g_m$, (b) subthreshold swing $S$, (c) turn-on transition voltage $\Delta V_T$, (d) on-current for the constant current defined threshold, $I_{on}(CC)$, and (e) drain-induced barrier lowering, DIBL, with source/drain geometry.

### 1. Transconductance, $g_m$

As shown in Fig. 2(a) and 3(a), Γ-InGaAs had by far the greatest $g_m$ for end injection. The small transport mass in the Γ-valley of Γ-InGaAs produces a high injection velocity, which, along with limited backscattering in the channel, more than overcomes any reduced carrier concentration in the channel due to the lower quantum (density of states) capacitance. In contrast, for MV-InGaAs, the limited density of states in the Γ-valley pushes the carriers high into that valley, while quantum mechanical confinement substantially reduces the band offsets between the low density-of-states Γ-valley and high density-of-states L-valleys. Now, more readily than in the bulk considered previously, electrons transfer to L-valleys, which have much slower carriers and much higher scattering rates. As a result, $g_m$ is reduced not only as compared to that of Γ-InGaAs, but also as compared to Si devices in these simulations (analogous to reduction of the high-field electron velocity in bulk GaAs below that in bulk Si).

For the currently preferred wrapped saddle/slot and RSD contact geometries as modeled, the advantage of Γ-InGaAs $g_m$ continues, but both the absolute value and the degree of advantage of Γ-InGaAs over the other systems in $g_m$ decreases substantially, as shown in Figs. 2(b) and (c). Moreover, both the RSD and saddle contacts somewhat favor a ⟨110⟩ channel orientation for Si, while, if anything, end contacts slightly favor a ⟨100⟩ channel orientation, which suggests that the ⟨110⟩ channel orientation advantage for RSD and saddle contacts is associated with contact geometry and not transport through the quantum-confined channel.

As noted previously, there are limitations to the comparison between Si and InGaAs devices, including using the same surface roughness (but, again, not surface roughness scattering rates) for all devices, and limitations in the band structure models, including the treatment of doping in the source and drain and the peripheral valley energy offsets in InGaAs. Perhaps more reliable are the comparisons among different device contact geometries for devices with the same material and channel orientation assumptions. Notably, returning to Fig. 3(a), the performance as a function of contact geometry in terms of transconductance, $g_m$, degrades for all material and channel orientation when going from the end-injection contact geometry to the saddle/slot and RSD contact geometries for ideal contacts, but least so for ⟨110⟩ channel orientation Si, and most so for Γ-InGaAs. Consistent with that latter result, while Γ-InGaAs maintains an advantage over MV-InGaAs for all contact geometries, that advantage is by far greatest for end-injection.

Toward understanding these dependencies of transconductance $g_m$ on contact geometry, compare the qualitative expectations for illustrative purely diffusive and purely ballistic transport models. For a diffusive model, for each considered material system, the transconductance $g_m$ should be greatest for the saddle geometry, and worst for the end-contact geometry, because of the proximity of the S/D contacts to the channel and contact surface area.

However, as the ballistic limit of transport is approached, contact alignment with the channel becomes increasingly important. Such a dependence was noted years ago in SCMC simulations of nano-scale planar III-V MOSFETs, where it was shown that poor coupling into the channel for substantially ballistic carries injected from surface contacts led to substantial "source starvation" in III-V systems and associated limits on transconductance [4]. To better understand the effects of contact alignment in the results of Fig. 3(a) as the ballistic limit is approached in the device geometries considered here, again note the assumptions of ideal perfectly injecting and absorbing contacts and equal-angle-reflecting closed boundaries used in simulations so far. To those assumptions add a few more chosen for illustrative value (only) in the immediate discussion here: drain voltages sufficiently large that electron injection from the drain to source can be neglected; a unity transmission probability for electrons reaching the source extension with sufficient kinetic energy along the channel to overcome the channel potential barrier, and a zero transmission probability otherwise, which makes the former the only electrons of interest here and the source extension a perfectly absorbing boundary for these electrons of interest; and a uniform (i.e., a perfectly-screened) potential within the source region. Ballistic ray tracing in this system shows that the saddle/slot contact MOSFET with a reflecting boundary at the end of the source region located at $L_C$ from the edge of the source extension boundary (Fig. 1), may be replaced by two mirror image MOSFETs with a saddle/slot contact around a common source region of length $2L_C$ connected to the drain extensions of both MOSFETs, without affecting injection of the electrons of interest into the drain extension and channel beyond. Similarly, without affecting injection of the electrons of interest into the drain extension, the end contact MOSFET with source length $L_C$ may be replaced by one with source length $2L_C$ (or of any other length). In turn, again without affecting the injection of the electrons of interest into the drain extension, this latter end contact MOSFET with source length $2L_C$, may be replaced by a MOSFET with a source length of $2L_C$ with both an end contact and a saddle/slot contact. (After this latter step, there would be considerably more backward moving carriers with enough energy to overcome the channel barrier originating from the side contacts, but these electrons can never return to the drain extension under these assumptions.) Thus, in the ballistic limit under these assumptions, the difference between the here-considered end contact MOSFET and saddle/slot contact MOSFET corresponds to the difference between having a both injecting and absorbing contact, or just an absorbing contact, respectively, at the end of a source region of length $2L_C$, with an injecting and absorbing saddle/slot contact about the source sides and top in either case. That difference is enhanced by electron injection probabilities that are peaked naturally about the surface normal direction, which selects for surfaces that are aligned perpendicular to the drain extension entry, as illustrated in Fig. 4 for injection about the end-contact-normal plane running along the vertical plane of Si semiconductor fins. We also note that a ray-tracing analysis under these conditions for the raised source drain geometry gives the same results as for the saddle geometry.

Moreover, ⟨100⟩ channel orientations have {100} contact surfaces, which promote a still greater peaking of the electron injection about the surface normal (plane), as also shown in Fig. 4. This peaking means that the loss of the injecting end contact of the hypothetical effective $2L_C$ source length saddle/slot MOSFET as discussed above, is more significant for the simulated Si ⟨100⟩ devices as compared to the Si ⟨110⟩ devices, which provides an advantage for the Si ⟨110⟩ device, consistent with the full simulation results of Fig, 3(a).

In the full simulations of Fig. 3(a), which contact geometry would be best and to what degree depends on scattering too, as well as other non-idealities of course. As measured by the larger transconductances for the reference end contacts, as show in Fig. 3(a), results for all of the considered materials systems would appear to lean toward the expectations of ballistic transport in the here-modeled nano-scale MOSFETs, greatly so for Γ-InGaAs, but only marginally so for the remaining systems with slower carriers and greater scattering rates.

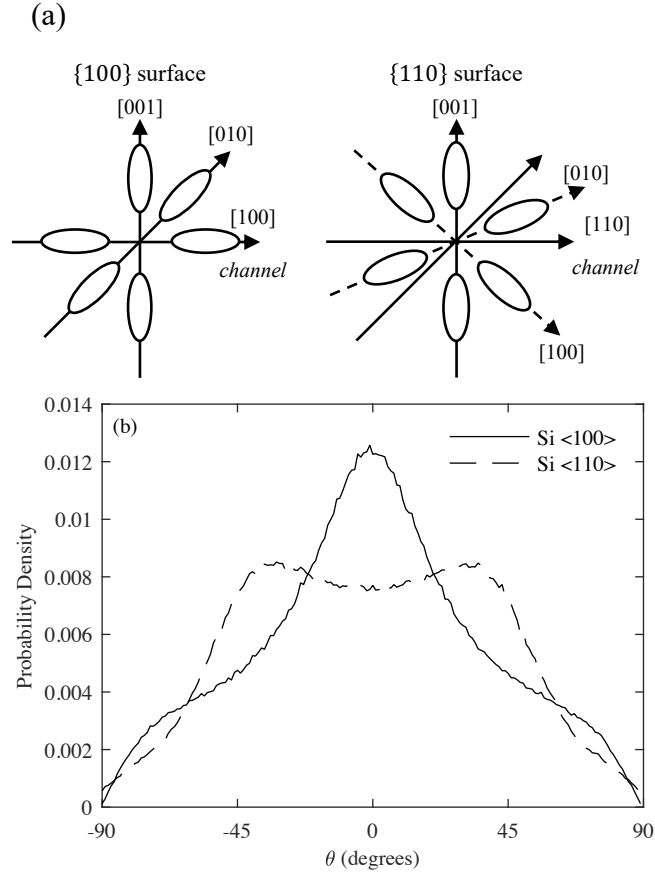

**FIG. 4.** (a) Alignment of the conduction channel relative the Si conduction band energy valleys for (on the left) ⟨100⟩ and (on the right) ⟨110⟩ channel orientations on a {100} substrate. (b) UTMC-simulated probability density per degree of the carrier injection angle with respect to the plane of the channel for channel-end-injected carriers, for Si ⟨100⟩ (solid line) and ⟨110⟩ (dashed line) channel orientations.

### *2. S, DIBL, turn-on characteristic, and on-state current*

In terms of electrostatic control, most notably, the Si devices also have better (smaller) off-state subthreshold swing $S$ and, more so, DIBL—where the latter is measured with a narrower potential barrier along the channel—than InGaAs devices (Figs. 3(b) and (e), respectively). The reason for this difference between Si and InGaAs devices is most likely the difference in dielectric constants leading to stronger coupling for the latter to the source and drain regions. Moreover, this difference is most significant for the saddle/slot geometry with the potential pinned at the outer edges of the separate confinement regions, and the least significant for end contacts where coupling to the source and drain is weaker as the band bending extends a few screening lengths into the source and drain.

Turn-on behavior, as measured by $\Delta V_T$, (Fig. 3(c)) is the slowest for Γ-InGaAs and, unlike for $S$ and DIBL, also differs substantially from that of MV-InGaAs. These differences suggest that it may be related to the smaller quantum capacitance of the Γ-InGaAs as compared to the Si and even to MV-InGaAs devices. As a result of this slow turn-on characteristic, Γ-InGaAs has a lower $I_\text{on}$ with a constant-current-defined threshold for the saddle/slot and RSD contact geometries compared to Si, despite better peak $g_\text{m}$. Γ-InGaAs, still provides an $I_\text{on}$ advantage for end contacts, but considerably less than for $g_\text{m}$.

### *3. Drain current vs. drain voltage*

The drain current also was calculated at the overdrive gate voltage of $V_\text{GS} - V_\text{T}^\text{CC} = 0.35$ V as a function of drain voltage $V_\text{DS}$ swept from 0 V to 0.6 V in steps of 25 mV, consistent with the transistor in the on-state with $V_\text{T}^\text{CC} = 0.25$ V and a $V_\text{DD} = 0.6$ V, as shown in Fig. 5. All devices with all contact configurations showed onset of saturation at between approximately $V_\text{DS} = 0.20$ V and 0.25 V, except for the Γ-InGaAs with end contacts, where the onset was delayed approximately 0.05 V. The saturation, however, was perhaps slightly better—less dependence on $V_\text{DD}$—for both of the InGaAs materials systems for all contact geometries.

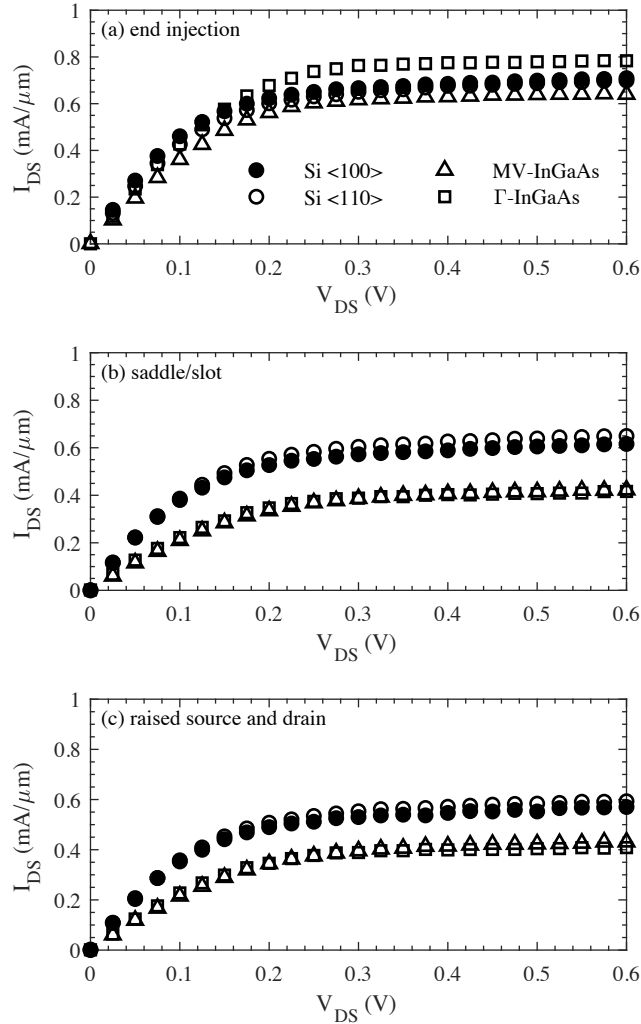

**FIG. 5.** $I_\text{DS}$-$V_\text{DS}$ simulation results for 18 nm Si ⟨110⟩ (open circles), Si ⟨100⟩ (solid circles), MV-In$_{0.53}$Ga$_{0.47}$As (open triangles), and Γ-In$_{0.53}$Ga$_{0.47}$As (open squares) FinFETs at the gate overdrive voltage of 0.35 V above the constant current threshold voltage for (a) end injection, (b) saddle/slot, and (c) raised source and drain.

**B. Non-Ideal Contacts**

In this section, the impact of parasitic (less than-ideal) S/D contact resistance on the performance of the considered FinFETs is examined. We re-consider only the transconductance, $g_m$, turn-on characteristic, $\Delta V_T$, on state current, $I_{on}$, and the $I_{DS}$ vs. $V_{DS}$ characteristic. Subthreshold swing $S$ and drain-induced barrier lowering $DIBL$ are calculated in the absence of any significant current flow and, thus, contact ideality is moot.

A typical approach to incorporate the effects of contact interface resistance due to imperfect coupling among the metal, silicide, and semiconductor is to apply a lumped resistance model as a post-processing step to the intrinsic $I_{DS}$-$V_{GS}$ characteristics using a self-consistent voltage-drop algorithm [31]. Here, however, we model non-ideal contacts directly in the Monte Carlo simulations via a reduction below unity in the probability for an electron to be transmitted across the contact interface in either direction, $T$. (Equal angle reflection is used to model carriers reaching, but not being transmitted across the contact interface from the inside.) In this way we preserve any contact geometry and surface orientation effects beyond just total surface area and avoid the need for such post-processing. The resulting apparent specific contact resistivity is, $\rho_{sp} = \rho_{LB}(T^{-1} - 1/2)$, where $\rho_{LB}$ is the Landauer-Büttiker ballistic resistivity [18,19].

In this work, for all considered devices, we employ an illustrative fixed value of $T = 0.2$ as a control, which produces $\rho_{sp} = 4.5\rho_{LB}$. For silicon, $\rho_{LB} = 3.0\times10^{-10}$ $\Omega$-cm$^2$ at the considered $2.0\times10^{20}$ cm$^{-3}$ doping concentration. Therefore, the corresponding specific contact resistivity is, $\rho_{sp} = 1.35\times10^{-9}$ $\Omega$-cm$^2$, which is reasonably near a state-of-the-art reported value of $1.2\times10^{-9}$ $\Omega$-cm$^2$ [32]. For the Γ-InGaAs and MV-InGaAs devices with $\rho_{LB}$ values of $1.3\times10^{-9}$ $\Omega$-cm$^2$ and $1.4\times10^{-9}$, respectively, at the considered $5.0\times10^{19}$ cm$^{-3}$ doping concentration, this control value of $T$ results in substantially larger $\rho_{sp}$ values, of $5.9\times10^{-9}$ $\Omega$-cm$^2$ and $6.3\times10^{-9}$ $\Omega$-cm$^2$, respectively, still somewhat better than reported values of $7\times10^{-9}$ $\Omega$-cm$^2$ [33] for InGaAs, which requires more careful materials processing than Si to develop ohmic contacts [34].

### *1. $g_m$, $\Delta V_T$, and $I_{on}$*

Overall, Fig. 6 shows that, as expected, non-ideal contacts decrease the peak transconductances and on-currents, and, with the RSD and the model saddle/slot contact geometries having approximately 3.3 and 2.5 times the contact surface area as the end contact geometry, the relative reduction is the greatest for the end-contacts. The RSD geometry to some degree has greatest $g_m$ for all materials systems. However, the saddle/slot geometry produces an $I_{on}$ comparable to that of the RSD geometry for silicon ⟨110⟩ channel devices, and greater than that of the RSD geometry for silicon ⟨100⟩ channels. Moreover, also as expected, detrimental effects are the greatest on the Γ-InGaAs devices followed by the MV-InGaAs devices. All Si devices with all contact geometries now outperform all of their InGaAs counterparts in terms of transconductance $g_m$ and, more so due to the slower turn-on characteristic for the latter, on-current with respect to a constant current threshold $I_{on}^{(CC)}$.

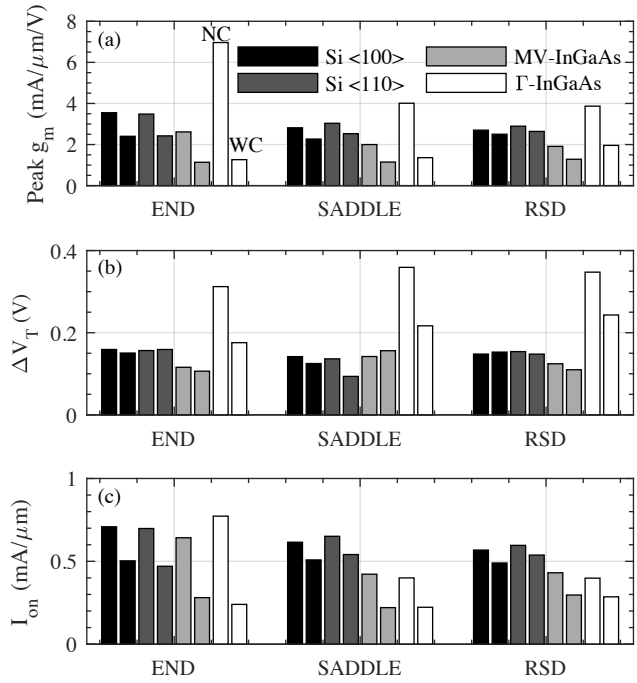

**FIG. 6.** Comparison of ideal and non-ideal contacts (a) on the peak transconductance $g_m$, (b) turn-on transition voltage $\Delta V_T$, and (c) on the on-current $I_{on}$ with a constant current defined threshold ($I_{on}$(CC)), for the end, saddle/slot, and RSD contacts to a 18 nm channel FinFETs at $V_{DS}$ of 0.6 V. Here two bars corresponding to with no added contact resistivity (NC) and with added contact resistivity (WC), respectively, are shown side by side on the same gray scale for each considered material system, including channel orientation for Si.

Despite the non-ideal contacts, there remains a notable and essentially undiminished advantage for the silicon ⟨110⟩ channel saddle/slot and RSD contact devices over their silicon ⟨100⟩ counterparts. However, for the purpose of illustration, consider that in the limit of very small transmission probabilities, essentially any electron lucky enough to be injected into the source region across any point along the contact surface at any angle ultimately will make it to the channel, so the preceding ray-tracing arguments alone are not sufficient. However, on average, each carrier injected through the contacts of one of these saddle/slot or RSD ⟨100⟩ devices, injected closer to normal to the contact interfaces on average than their ⟨110⟩ surface devices, now will take slightly longer to reach the channel, creating a greater time-weighted charge contribution within the source region. Thus, self-consistent electrostatics now may compensate for reduced interface transmission probabilities by adjusting the source potential profile somewhat to effectively reduce the relative overall injection of electrons into the source of the saddle/slot or RSD ⟨100⟩ devices.

### *2. Drain current vs. drain voltage, revisited*

Fig. 7 shows drain current $I_{DS}$ vs. drain voltage $V_{DS}$ for the considered silicon and InGaAs devices. Performance degradation consistent with the $I_{on}$ of Fig. 6 is evident. However, there is notable lack of significant stretch-out in the $I_{DS}$-$V_{DS}$ characteristics relative to those for ideal contacts, which is perhaps most notable for end contact Γ-InGaAs, which, by far, shows the most reduction in $g_m$. If the contact resistances were post-processed, a substantial stretch-out in $I_{DS}$-$V_{DS}$ would be expected for this device, with much of $V_{DS}$ dropped across the contact resistances instead of between them. However, transport is near ballistic for the simulated Γ-InGaAs device (as previously discussed), and the voltage between the contacts is, thus, not well defined. In particular, to the extent transport between the contact and channel is ballistic, the shape of the energy distribution of injected carriers reaching the channel remains the same; there are just fewer of them. Contact post-processing represents (among other things) the opposite limit, where there is a well-defined voltage drop between the contacts and the channel consistent with charge carriers that are well thermalized to the crystal lattice when entering (leaving) the source (drain). The silicon devices in this work show less, if still significant, ballistic transport (also as previously discussed) so that a post-processing model would be more reliable at least, but the contact resistivity also is less important.

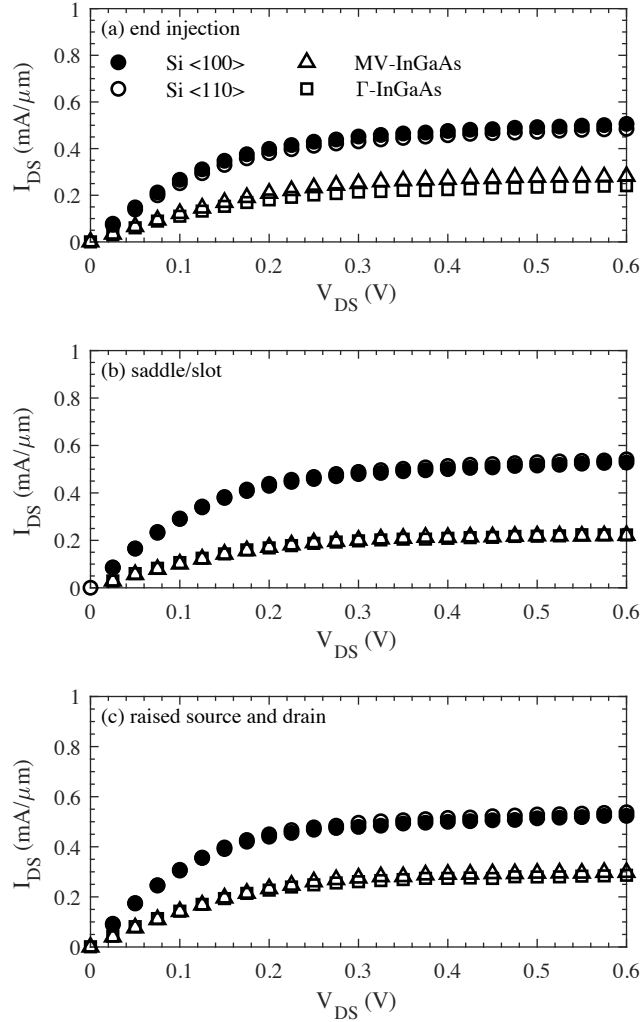

**FIG. 7.** As for Fig, 5 but with non-ideal contacts, $I_{DS}$-$V_{DS}$ simulation results for 18 nm Si ⟨110⟩ (open circles), Si ⟨100⟩ (solid circles), MV-In$_{0.53}$Ga$_{0.47}$As (open triangles), and Γ-In$_{0.53}$Ga$_{0.47}$As (open squares) FinFETs at the overdrive gate voltage of 0.35 V above the constant current threshold voltage for (a) end injection, (b) saddle/slot (where the MV-InGaAs and Γ-InGaAs data are difficult to distinguish), and (c) raised source and drain.

## IV. CONCLUSION

The effects of contact geometry and contact ideality on InGaAs and Si nano-scale n-channel FinFET performance are studied using a quantum-corrected semi-classical Monte Carlo method. End, saddle/slot, and raised source/drain contacts were modeled, and with ideal contact transmissivity and with reduced transmissivity more consistent with experimental contact resistivities. Far-from-equilibrium degenerate statistics, quantum-confinement effects on carrier distributions in real-space and among energy valleys, quasi-ballistic transport inaccessible through drift-diffusion and hydrodynamic simulations, and scattering mechanisms and contact geometries not readily accessible through NEGF simulation are addressed. Si ⟨110⟩ channel devices, Si ⟨100⟩ channel devices, multi-valley (MV) InGaAs devices with conventionally-reported energy valley offsets, and idealized Γ-valley only (Γ) InGaAs devices are modeled. Simulated silicon devices exhibited relatively limited degradation in performance due to non-ideal contact transmissivities, more limited sensitivity to contact geometry with non-ideal contact transmissivities, and some contact-related advantage for Si ⟨110⟩ channel devices. In contrast, simulated InGaAs devices were highly sensitive to contact geometry and ideality and the peripheral valley energy offset. For example, despite challenges to simulation-based comparison with different channel materials, it is illustrative of this latter sensitivity that simulated Γ-InGaAs device outperformed all others by a factor of two or more in terms of peak transconductance with perfectly transmitting reference end contacts, although significantly less so in terms of on state current above a constant current threshold due to a poorer turn on characteristic, while silicon devices outperformed Γ-InGaAs for all contact geometries with the considered non-ideal contact resistivities, and MV-InGaA devices performed the poorest under all simulation scenarios.

In summary, the substantial advantage of the simulated ideal end-contact Γ-InGaAs device, at least in transconductance, continues to bear out the motivation for the use of such high-thermal-velocity materials. However, common contact geometries, currently reasonable contact resistivities, and substantial peripheral valley occupation, represent performance bottlenecks that can degrade and more than eliminate that otherwise expected advantage, but perhaps also represent opportunities for improvement.

**ACKNOWLEDGMENTS**

This work was supported by a Cockrell School of Engineering at The University of Texas at Austin graduate fellowship. We thank the Texas Advanced Computing Center (TACC) for generous supercomputing resources and software consultation. S. K. Banerjee was supported by the NSF NASCENT ERC, and the NSF NNCI grant 1542159.